\documentclass[conference]{IEEEtran}
\IEEEoverridecommandlockouts
\usepackage{cite}
\usepackage{amsmath,amssymb,amsfonts}
\usepackage{algorithmic}
\usepackage{graphicx}
\usepackage{textcomp}
\usepackage{xcolor}
\usepackage{marvosym}
\usepackage[normalem]{ulem}
\usepackage{multirow}
\usepackage{algorithm}
\usepackage{algorithmic}
\useunder{\uline}{\ul}{}
\def\BibTeX{{\rm B\kern-.05em{\sc i\kern-.025em b}\kern-.08em
    T\kern-.1667em\lower.7ex\hbox{E}\kern-.125emX}}
\begin{document}

\title{Enhancing Job Recommendation through LLM-based Generative Adversarial Networks
}

\author{\IEEEauthorblockN{Yingpeng Du\textsuperscript{1,*,\Letter}, Di Luo\textsuperscript{2,*}, Rui Yan\textsuperscript{2,\Letter}, Hongzhi Liu\textsuperscript{1}, Yang Song\textsuperscript{3,\Letter}, Hengshu Zhu\textsuperscript{4}, Jie Zhang\textsuperscript{5}
\thanks{\textsuperscript{*} Equal Contribution.}
\thanks{\textsuperscript{\Letter} Corresponding authors.}}
\IEEEauthorblockA{
\textit{$^{1}$School of Software and Microelectronics, Peking University}.\\
\textit{$^{2}$Gaoling school of Artificial Intelligence, Renmin University of China},\\
\textit{$^{34}$NLP Center, Career Science Lab, BOSS Zhipin}.\\
\textit{$^{5}$School of Computer Science and Engineering, Nanyang Technological University}.\\
dyp1993@pku.edu.cn, di\_luo@ruc.edu.cn, ruiyan@ruc.edu.cn, liuhz@pku.edu.cn, \\
songyang@kanzhun.com, zhuhengshu@gmail.com, zhangj@ntu.edu.sg}}


\maketitle

\begin{abstract}
Recommending suitable jobs to users is a critical task in online recruitment platforms, as it can enhance users' satisfaction and the platforms' profitability. While existing job recommendation methods encounter challenges such as the low quality of users' resumes, which hampers their accuracy and practical effectiveness.
With the rapid development of large language models (LLMs), utilizing the rich external knowledge encapsulated within them, as well as their powerful capabilities of text processing and reasoning, is a promising way to complete users' resumes for more accurate recommendations. However, directly leveraging LLMs to enhance recommendation results is not a one-size-fits-all solution, as LLMs may suffer from fabricated generation and few-shot problems, which degrade the quality of resume completion.

In this paper, we propose a novel LLM-based approach for job recommendation. To alleviate the limitation of fabricated generation for LLMs, we extract accurate and valuable information beyond users' self-description, which helps the LLMs better profile users for resume completion. Specifically,  we not only extract users' explicit properties (e.g., skills, interests) from their self-description but also infer users' implicit characteristics from their behaviors for more accurate and meaningful resume completion.
Nevertheless, some users still suffer from few-shot problems, which arise due to scarce interaction records, leading to limited guidance for the models in generating high-quality resumes.
To address this issue, we propose aligning unpaired low-quality with high-quality generated resumes by Generative Adversarial Networks (GANs), which can refine the resume representations for better recommendation results.
Extensive experiments on three large real-world recruitment datasets demonstrate the effectiveness of our proposed method.

\end{abstract}


\section{Introduction}

\begin{figure*} \centering
 \includegraphics[width=1.0\textwidth]{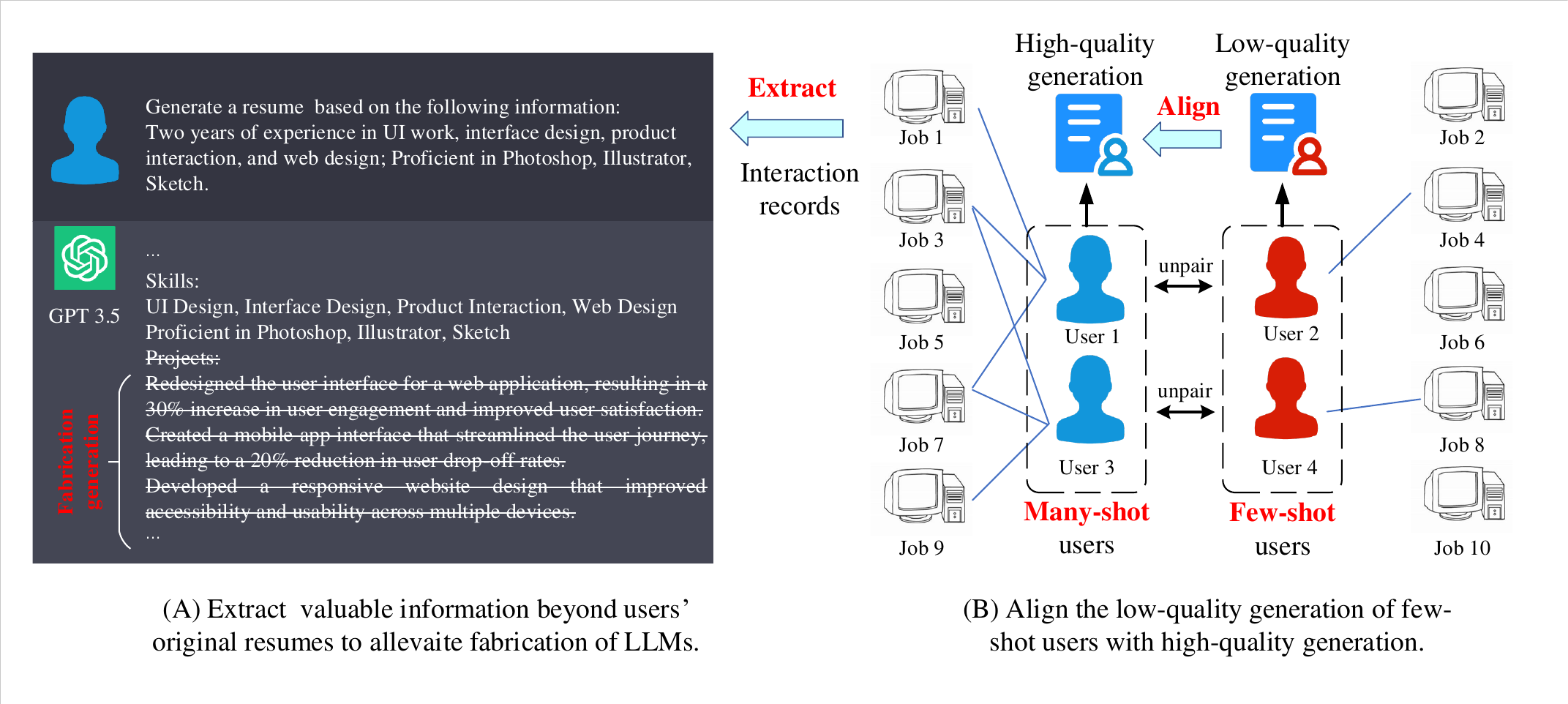}
\caption{The difficulty and motivation behind leveraging LLMs for job recommendation.} \label{intro}
\end{figure*}

Job recommendation is the most fundamental and essential task of today's online recruitment platforms as it serves as the cornerstone for efficient recruitment processes. It can greatly improve recruitment efficiency by accurately providing job seekers (also known as users) with suitable job positions. Although existing job recommendation methods have achieved considerable success in recent years, they still face significant challenges, such as the low quality of users' resumes and interference from few-shot problems, which hinder their accuracy and efficiency in practical applications.
For example, some users may not invest sufficient effort in crafting their resumes, resulting in difficulties in understanding their skill abilities and job preferences. Other users may lack a comprehensive self-awareness of their own skill abilities and preferences, further diminishing the quality of their resumes.
Inspired by the recent remarkable capabilities and rapid development of large language models (LLMs), it is intuitive to utilize the vast external knowledge stored in LLMs, as well as their powerful text comprehension and completion capabilities, to improve and rectify low-quality resume descriptions. This can bridge the information gaps between user resumes and job descriptions for better recommendation.

However, simply leveraging LLMs to enhance the quality of users' resumes is not a one-size-fits-all solution for job recommendation. Due to the widespread fabrications and hallucinations that occur within LLMs, it is challenging to generate high-quality resumes for users unless reliable interactive information and appropriate generated data alignment are provided. Fig.\ref{intro} (A) illustrates the resume generation process for a user using simple completion with LLMs. It highlights that the generated results often contain massive amounts of unrelated and fabricated information, rendering them unsuitable for effective recommendation. To alleviate the limitation of fabricated generation in LLMs, we propose exploring users' interactive behaviors (e.g., reach an agreement for an interview) with recommender systems to mine their relevance to users' abilities and preferences, thereby assisting the LLMs in better profiling users for resume completion. For instance, users typically possess specific job skills, residential addresses, and educational backgrounds, which influence their interactions with job positions that have corresponding job responsibilities, locations, and levels. Consequently, we propose inferring users' implicit characteristics (e.g., skills, interests) from their interaction behaviors, as this can enhance the LLMs' ability to accurately profile users and generate meaningful resumes.

Although exploring users' interactive behaviors contributes to better profiling users by LLMs for resume completion, it may still suffer from the few-shot problems, limiting the quality of resume completion for certain users. On the one hand, users with rich interaction records enable us to generate high-quality resumes based on them. On the other hand, users with few interaction records widely existed in real-world scenarios (also known as the long-tail effect) face challenges with the fabrications and hallucinations of LLMs, as they lack sufficient interactive information. To address this problem, we propose aligning the generated resumes of few-shot users with the high-quality resumes of users who have extensive interaction records. To align the unpaired resumes across different users, we introduce a transfer learning method based on generative adversarial networks (GANs) to refine the generated resumes of few-shot users for job recommendation.
Specifically, the generator in GANs aims to improve the representations of low-quality resumes by fooling the discriminator. Meanwhile the discriminator in GANs aims to distinguish between the refined representations and the high-quality representations as much as possible.
Through iterative training of GANs, the generator plays a crucial role in refining the representations of low-quality resumes, leading to more accurate recommendation results. By leveraging the power of GANs, we can bridge the gap between few-shot users and users with rich interaction records, thereby enhancing the quality of resume completion for all users.

In this paper, we propose an LLM-based GANs Interactive Recommendation (LGIR) method for job recommendation. The goal is to address the limitations of fabricated generation in LLMs and the few-shot problems that affect the quality of resume generation.
To tackle the LLMs' fabricated generation limitation, we extract accurate and valuable information beyond users' submitted resumes, i.e., not only extracting users' explicit properties (e.g., skills, interests) from their self-description but also inferring users’ implicit characteristics from their behaviors for more accurate and meaningful resume completion.
To address the few-shot problems, which restrict the quality of generated resumes, we propose the transfer representation learning for low-quality resumes using GANs, which can align the low-quality resumes with unpaired high-quality resumes, leading to improved recommendation results.
By simultaneously integrating LLM-based interactive resume completion and aligning low-quality with high-quality resumes within a single framework, our proposed method accurately captures users' skills and preferences, thereby enhancing the effectiveness of job recommendation results.
We evaluate our model on three real-world datasets, and demonstrate its consistent superiority over state-of-the-art methods for job recommendation. Ablation experiments and a case study further validate the motivations behind our proposed method.

\section{Related Work}
\subsection{Job Recommendation}

Job recommendation has gained significant popularity in online recruitment platforms. It offers the potential to enhance recruitment efficiency by accurately matching job seekers with suitable job positions. Unlike conventional recommendation systems, job recommendation in online platforms benefits from a wealth of textual information, including users' resumes and job requirement descriptions. These textual cues provide valuable insights that can be leveraged to improve the accuracy and relevance of job recommendations. According to the way of using information for job recommendation, existing methods can be primarily categorized into three groups: behavior-based method, content-based method, and hybrid method.
Behavior-based methods have been developed to leverage user-item interaction for job recommendation.
Among the methods, collaborative filtering (CF) based methods \cite{su2009survey} gain popularity due to their good performance and elegant theory. Classical CF-based methods model users' preferences by exploiting their historical interactions, such as matrix factorization \cite{koren2009matrix}.
Recently,  augmenting CF techniques with delegate modification has become increasingly popular, including utilizing deep neural networks \cite{he2017neural,zhang2019deep,xue2017deep} and graph models such as GCNs \cite{he2020lightgcn,wang2019neural,wang2019kgat}.
Content-based methods have been developed to leverage the rich semantic information present in resumes and job requirements using text-matching strategies or text enhancement techniques. These methods include CNN \cite{zhu2018person, shen2018joint}, RNN \cite{qin2018enhancing}, and memory networks \cite{yan2019interview}.
For instance, APJFNN \cite{qin2018enhancing} introduced hierarchical attention RNN models for person-job fit, which captured word-level semantic representations for both user resumes and job descriptions. Additionally, various techniques have been explored to enhance the expressive power of text encoders. These include adversarial learning \cite{luo2019resumegan}, co-teaching mechanisms \cite{bian2020learning}, and transfer learning \cite{bian2019domain}.
The hybrid methods have been developed to leverage the rich textual information as well as the user-item interaction for job recommendation. Specifically, they construct the embeddings of users and jobs based on their text content, and leverage user-item interaction for job recommendation. For example, IPJF \cite{le2019towards} adopted a multi-task optimization approach to learning the perspectives/intentions of users and jobs based on shared text features. SHPJF \cite{hou2022leveraging} utilizes both text content from jobs/resumes and search histories of users for job recommendation. 
However, these methods either suffer from low quality of users' resumes or few-shot problems. To this end, we propose to utilize the rich external knowledge and generation capabilities encapsulated within LLMs to improve the quality of users' resumes for accurate recommendation results.

\subsection{Large Language Models for Recommendation}

With the popularity of Large Language Models (LLMs) \cite{touvron2023llama, brown2020language} in the field of Natural Language Processing (NLP), there is a surging interest in leveraging these advanced models to amplify the efficacy of recommendation systems \cite{wu2023survey}. These models, which are trained on voluminous data sets via self-supervision, possess the potential to boost the quality of recommendations due to their extensive assimilation of external knowledge \cite{liu2023pre}. This characteristic sets them apart from conventional recommendation systems: LLM-based models have the ability to capture contextual information with superior efficacy, enabling a precise understanding of user queries, product descriptions, and other forms of textual data \cite{geng2022recommendation}. The heightened comprehension of context by the LLMs consequently improves the accuracy and relevance of recommendations, ultimately leading to increased user satisfaction.

Moreover, facing the cold-start problem \cite{da2020recommendation}, LLMs also introduce innovative solutions to recommendation systems via their zero-shot recommendation capabilities \cite{sileo2022zero}. Due to comprehensive pre-training incorporating factual data, domain knowledge, and common-sense reasoning, these models can effectively generalize to previously unseen users, which enables them to provide plausible recommendations without prior exposure to specific items or users.

Additionally, LLMs have the potential to improve the interpretability of recommendations. They are capable of generating language-based explanations for their choices \cite{gao2023chat}. This not only allows people to comprehend the factors influencing the recommendations but also enhances trust in the system.

Despite its excellent capacities, directly leveraging the parametric knowledge stored in LLMs for generating recommended items faces several challenges \cite{liu2023chatgpt}. Firstly, there are knowledge gaps between LLMs and items that require recommendation, meaning that the LLMs might not possess sufficient information or understanding to generate accurate and relevant recommendations for them. Secondly, LLMs have a proclivity for producing incomplete and unrealistic results, necessitating an additional step to ground the recommendations in relevant knowledge and rule out undesirable outcomes.

In response to these issues, several recent studies have focused on controlling and directing the output of the LLMs through the constructive design of prompts and with the help of in-context learning, aiming to alleviate hallucinations and allow an efficient adaptation of the LLMs to downstream recommendation tasks without the need for tuning model parameters \cite{wu2023survey}. 
For instance, Hou et al. \cite{hou2023large} introduced two prompting methods to enhance the sequential recommendation capabilities of LLMs.
Gao et al. \cite{gao2023chat} designed an interactive recommendation framework centered around ChatGPT, leveraging multi-turn dialogues to understand user requirements and utilizing existing recommendation systems to get results. 
Wang et al. \cite{wang2023generative} proposed a generative recommendation framework that employs LLMs to determine when to recommend existing items or generate new items using AIGC models.

We also notice a few works attempt to explore the integration of user behavior history to serve as guidelines for LLMs \cite{chen2023palr}.
However, these recommendation methods are still not immune to pervasive long-tail issues in recommendation systems. These problems arise due to limited interaction records, providing scarce guidance and constraints to these models.
The lack of constraints can worsen what is known as the model's hallucination issue, where it tends to generate false or inaccurate texts, which in turn, can skew the recommendations provided by the system.
To combat this few-shot problem, we are among the first to use Generative Adversarial Networks (GANs) to align the low-quality generated resumes of these few-shot users with the high-quality resumes of users who have ample interaction records. This alignment process aims to enhance the representations of low-quality resumes, ultimately improving the recommendation quality.

\begin{figure*} \centering
 \includegraphics[width=1.0\textwidth]{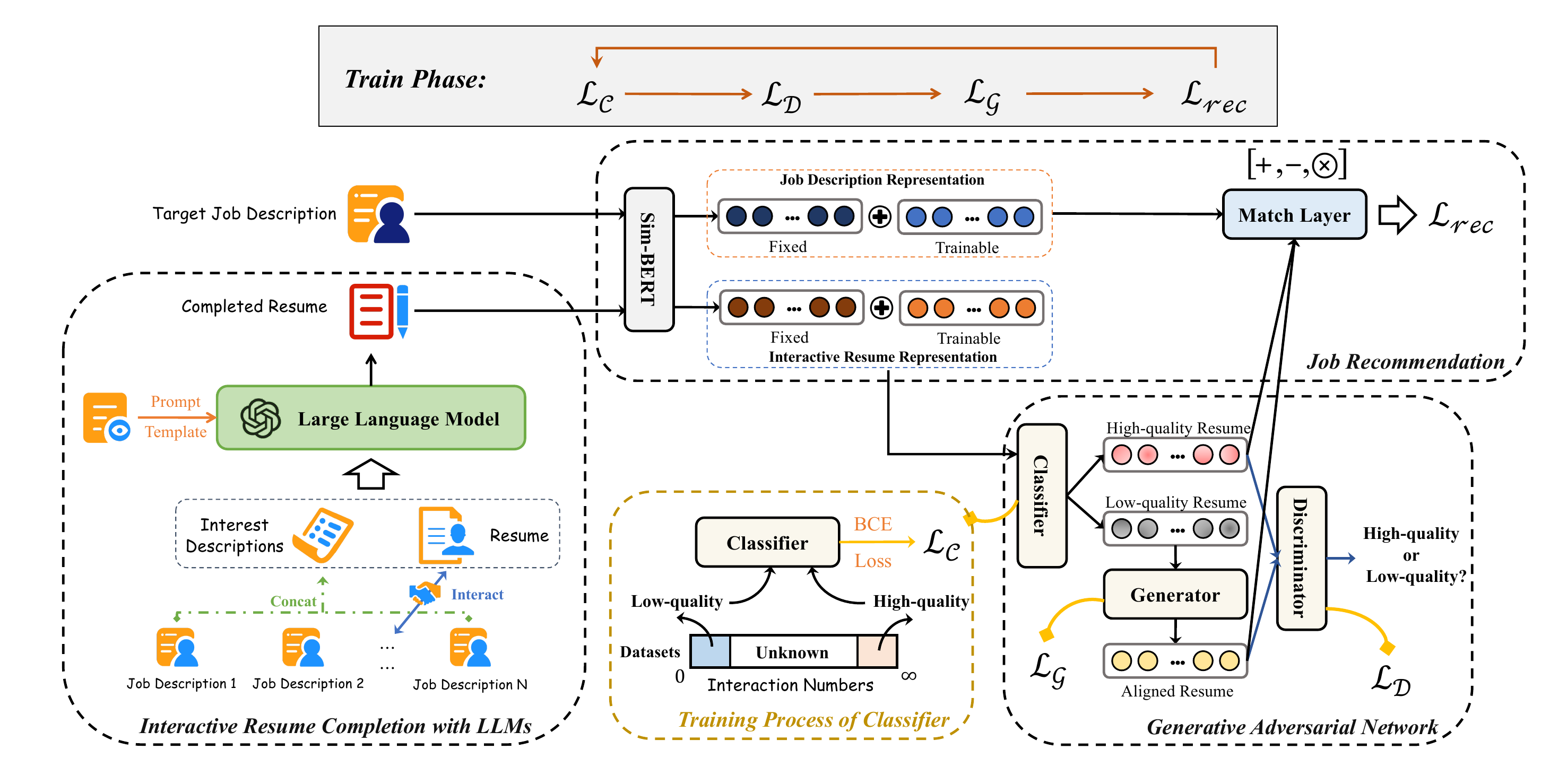}
\caption{The architecture of the LLM-based GANs Interactive Recommendation (LGIR), mainly contains the interactive resume completion method for resume generation and the GANs-based method for resume quality alignment.} \label{Method}
\end{figure*}

\section{Problem Definition}
Let $\mathcal{C}=\{c_1,\cdots,c_N\}$ and $\mathcal{J}=\{j_1,\cdots,j_M\}$ represent the sets of $N$ users and $M$ jobs, respectively. Each user or job has a text document describing the resume or job requirements of them. Specifically, we denote the resume of user $c$ as $T_c = [w_1,\cdots,w_{l_c}]$, where $w_i$
is the i-th word in the resume $T_c$ and $l_c$ denotes the 
the length of resume $T_c$. Similarly, we denote the requirement descriptions of job $j$ as $T_j = [w_1,\cdots,w_{l_j}]$, where $w_i$
is the i-th word in the resume $T_j$ and $l_j$ denotes the 
the length of requirements $T_j$. Besides the text document describing the resume or job requirements, we suppose to know the interaction records between users and jobs, which can be denoted as an interaction matrix $\mathcal{R}\in\mathbb{R}^{N\times M}$ as follows:
\begin{equation}
R_{ik}=\begin{cases} 1,  &\text{if user $u_i$ interacted with the job $j_k$};\\
0,  &\text{otherwise}  \\
\end{cases}
\end{equation}

In this paper, our goal is to recommend the appropriate
jobs to users. Formally, we propose learning a
matching function $g(c_i, j_k)$ based on the interaction records $\mathcal{R}$ and the documents describing the resume or job requirements, and then make the top-$K$ recommendation based on the matching function.

\section{The proposed method}

This paper focuses on how to make use of the rich external knowledge and the powerful text processing capabilities of LLMs for job recommendation, whose overall architecture is shown in Fig.(\ref{Method}).
First, we propose an interactive resume completion method to alleviate the limitation of the fabricated generation of LLMs. Second,  we propose a GANs-based method to refine the representations of low-quality resumes with alignment. Finally, we propose a multi-objective learning framework for job recommendation.

\subsection{A LLM-based Summary Method for Resume Completion}
To improve and rectify low-quality resume descriptions for job recommendation, we draw inspiration from the remarkable capabilities of LLMs to enhance the quality of users' resumes. By leveraging the extensive external knowledge stored within LLMs and their excellent abilities in understanding text and reasoning, we can explore promising avenues for improving the accuracy of job recommendation. In this section, we propose two kinds of ways, named Simple Resume Completion (SRC) and Interactive Resume Completion (IRC), to complete and improve the quality of users’ resumes for more accurate recommendation.

\begin{figure*} \centering
 \includegraphics[width=0.8\textwidth]{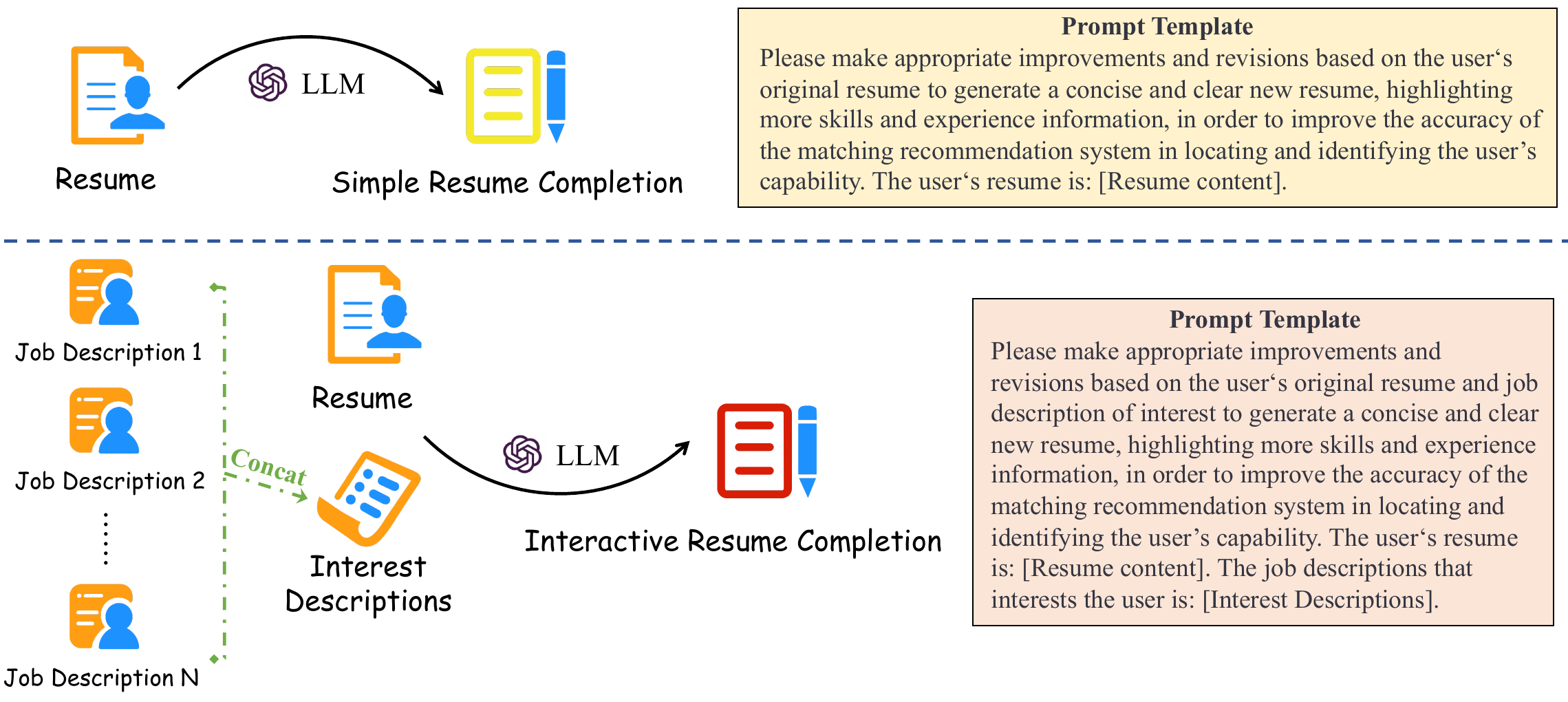}
\caption{The difference between Simple Resume Completion and Interactive Resume Completion, and how their prompts are designed.} \label{prompt}
\end{figure*}

\subsubsection{Simple Resume Completion with LLMs} 
To complete the resume of users, we propose a prompting approach by leveraging LLMs' knowledge and generation abilities. Specifically, we construct the prompt for LLMs to enrich resume information based on the user's self-description, i.e.,

\begin{equation}
G_c = \text{LLMs}(prompt_{\text{SRC}},T_c)
\end{equation}
where prompt denotes the command that triggers the LLMs to complete the user $u$'s resume based on his/her self-description $T_c$, whose detail is shown in the upper part of Fig. \ref{prompt}. However, the simple strategy may suffer from the limitation of the fabricated and hallucinated generation of LLMs.

\subsubsection{Interactive Resume Completion with LLMs} 
To alleviate the limitation of the fabricated generation in LLMs, we propose to exploring users' interactive behaviors with recommender systems, thereby assisting the LLMs in better profile users for resume completion. For instance, users typically possess specific job skills, residential addresses, and educational backgrounds, which influence their interactions with job positions that have corresponding job responsibilities, locations, and levels. Consequently, users' implicit characteristics (e.g., skills, interests) can be inferred from their interaction behaviors for more accurate and meaningful resume completion. Specifically, we adopt a specific prompting approach for resume completion by LLMs with consideration of both user's self-description and his/her interactive behaviors, i.e.,
\begin{equation}
G_c = \text{LLMs}(prompt_{\text{IRC}},T_c,R_c)
\end{equation}
where $R_c=\{T_{j_k}|\mathcal{R}_{c{j_k}}=1\}$ denotes the job requirements that
the user $c$ have interacted with. The detail of the prompt is shown in the lower part of Fig. \ref{prompt}.

To learn embeddings of users and jobs in the latent space, we first embed all users and jobs with two embedding matrices $P\in \mathbb{R}^{N\times d}$ and $Q\in \mathbb{R}^{M\times d}$, respectively. $d$ denotes the dimension of latent space. Then, we use the one-hot encoding to look up the embeddings $P_i\in \mathbb{R}^{d}$ and $Q_k\in \mathbb{R}^{d}$ of user $c_i$ and job position $j_k$,
\begin{equation*}
 P_i = \text{Onehot}(c_i)\cdot P,
\end{equation*}
\begin{equation*}
 Q_k = \text{Onehot}(j_k)\cdot Q,
\end{equation*}
where $\text{Onehot}(c_i) \in \mathbb{R}^{1\times N}$ and $\text{Onehot}(j_k)\in \mathbb{R}^{1\times M}$ denote the  one-hot encoding of user $c_i$ and job $j_k$, respectively.

To make use of the descriptive text which is associated with a user or job position, we adopt the BERT model to encode them into the constant text embedding  $W_{t}\in \mathbb{R}^{d}$  as in \cite{yang2022modeling}. Specifically, we first maintain the original text order and place a unique token $[CLS]$ before the text, then we feed the combined sequence into the SIM-BERT model and use the output of token $[CLS]$ as the semantic embeddings of the descriptive text. Finally, we concatenate the embeddings of users and jobs w.r.t. their text embeddings as the hybrid representations, i.e.,
\begin{equation}\label{eq14}
x_{c_i} = \text{MLP}_{\text{user}}([P_i;W_{G_{c_i}}]) ,
\end{equation}
\begin{equation}\label{eq15}
 x_{j_k} = \text{MLP}_{\text{job}}([Q_k;W_{T_{j_k}}]) ,
\end{equation}
where $G_{c_i}$ and $T_{j_k}$ denote the user $c_i$'s LLMs-generated resume and the job $j_k$'s requirement description. $\text{MLP}_{\text{user}}$ and $\text{MLP}_{\text{user}}$ denote the multi-layer perceptron with hidden layers $[ 2\cdot d\to d_{e'}\to d_{e}]$  and activation functions $\text{Relu}(\cdot) = \max(\cdot,0)$. $d_{e}$ and $ d_{e'}$ denote the dimension of hidden layers in the multi-layer perceptron.

\subsection{A GAN-based Aligning Method for generated resume}

Although exploring users' interactive behaviors help LLMs better profile users for resume completion, it may still suffer from the few-shot problems for most long-tail users, limiting the quality of generated resume in real-world scenarios. To this end, we propose a transfer representation learning to refine the low-quality resumes of few-show users, which mainly contains a classifier for low-quality resume detection and GANs for resume generation quality alignment.

\subsubsection{Classifier}
To detect the low-quality resumes for alignment, we propose a classifier $\mathcal{C}(x)$ to distinguish between high-quality resumes and low-quality resumes, i.e.,
\begin{equation}
 \mathcal{C}(x)  = \sigma(W^c_2\cdot \text{Relu}(W^c_1\cdot x))
\end{equation}
where $W^c_1\in \mathbb{R}^{d_{c}\times d_{e}}$ and $W^c_2\in \mathbb{R}^{1\times d_{c}}$ denotes the parameters in the classifier $\mathcal{C}$, and we assemble them as a set $\Theta_{\mathcal{C}} =\{W^c_1,W^c_2\}$. We assume some users with extremely few and rich interaction records generate the low-quality and high-quality resumes by LLMs, respectively. Therefore, we can learn from partial users to generalize for all users with the classifier $\mathcal{C}$. Specifically, we
introduce the cross-entropy loss to train the classifier $\mathcal{C}$ on these partial users to as follows:
\begin{equation}\label{loss_c}
\mathcal{L}_\mathcal{C}  = \mathbb{E}_{(c_i,y_{c_i})\sim T_\mathcal{C}}[ y_{c_i} \cdot \log( \hat{y}_{c_i}) + (1- y_{c_i}) \cdot \log(1- \hat{y}_{c_i})]
\end{equation}
where $\hat{y}_{c_i} = \mathcal{C}(x_{c_i})$ denotes the quality prediction for the generated resume of the user $c_i$ based on the classifier $\mathcal{C}$. $T_\mathcal{C} = T_\mathcal{C}^{\uparrow}\bigcup T_\mathcal{C}^{\downarrow}$ denotes selected user set for the classifier $\mathcal{C}$ training, where $T_\mathcal{C}^{\uparrow} = \{(c_i,1)| \sum_k R_{ik}\ge \kappa_1\}$ and $T_\mathcal{C}^{\downarrow} = \{(c_i,0)| \sum_k R_{ik}\le \kappa_2\}$ denotes the many-shot users and the few-shot users, $\kappa_1$ and $\kappa_2$ are set as the threshold to select the few-shot users and  many-shot users for classifier learning.  $y_{c_i}$ denotes the ground truth label where $y_{c_i} = 1$ if $c_i \in T_\mathcal{C}^{\uparrow}$ else $y_{c_i} = 0$ if $c_i \in T_\mathcal{C}^{\downarrow}$.

\subsubsection{Discriminator}
The main idea of the discriminator in GANs is to learn and distinguish between samples that are generated by the model (generated samples) and sampled from the real data distribution (real samples) \cite{goodfellow2020generative}. The discriminator is trained to assign a high probability to real samples and a low probability to generated samples.

In this paper, we propose a discriminator to determine whether a given resume representation is a result of the generator's refinement process or a direct encoding of a high-quality resume, i.e.,
\begin{equation}
 \mathcal{D}(x)  = \sigma(W^d_2\cdot \text{Relu}(W^d_1\cdot x))
\end{equation}
where $W^d_1\in \mathbb{R}^{d_{s}\times d_{e}}$ and $W^d_2\in \mathbb{R}^{1\times d_{s}}$ denote the parameters of discriminator $\mathcal{D}$, and we assemble them as a set $\Theta_{\mathcal{D}} =\{W^d_1,W^d_2\}$.

\subsubsection{Generator}
Correspondingly, we propose a generator $\mathcal{G}$ to refine the representations of low-quality resumes which is recognized by the proposed classifier.
Specifically, the generator $\mathcal{G}$ aims to map the low-quality representations of resumes to their high-quality representations for transfer learning, i.e.,
\begin{equation}
 \mathcal{G}(x)  = W^g_2\cdot \text{Relu}(W^g_1\cdot x)
\end{equation}
where $W^g_1\in \mathbb{R}^{d_{g}\times d_{e}}$ and $W^d_2\in \mathbb{R}^{d_{g}\times d_{e}}$ denote the parameters in the generator $\mathcal{G}$, and we assemble them as a set $\Theta_{\mathcal{G}} =\{W^g_1,W^g_2\}$.

During training, the generator generates samples that are then fed into the discriminator. The discriminator evaluates these generated samples and provides feedback to the generator on their quality. This feedback helps the generator improve its ability to produce samples that resemble the real data distribution more closely.

\subsubsection{Adversarial Learning}
To align representations of the low-quality and high-quality  resumes, we propose to conduct a mini-max game between a generator and a discriminator in GANs. Specifically, the discriminator's role is to distinguish whether a given resume representation is a result of the generator's refinement process or a direct encoding of a high-quality generated resume.
To train the discriminator model $\mathcal{D}$, we propose to maximize the probability of assigning the correct label to both high-quality resume representations and refined resume representations by generator model $\mathcal{G}$, i.e.,
\begin{equation}\label{loss_d}
 \max_{\Theta_{\mathcal{D}}} \mathcal{L}_\mathcal{D}  = \mathbb{E}_{{c_{i_1}}\sim \hat{T}_\mathcal{C}^{\uparrow} }[\log \mathcal{D}(x_1)] + \mathbb{E}_{c_{i_2} \sim \hat{T}_\mathcal{C}^{\downarrow} }[ 1 - \log \mathcal{D}(\mathcal{G}(x_{c_{i_2}}))]
\end{equation}
where $\hat{T}_\mathcal{C}^{\uparrow}=\{{c_{i_1}}|\mathcal{C}(x_{c_{i_1}})\ge 0.5 \land c_{i_1}\in {T}_\mathcal{C}^{\uparrow}\}$ and $\hat{T}_\mathcal{C}^{\downarrow} = \{{c_{i_2}}|\mathcal{C}(x_{c_{i_2}})< 0.5 \land c_{i_2} \in {T}_\mathcal{C}^{\downarrow}\}$ denote high-quality and low-quality generated resumes detected by the classifier $\mathcal{C}$.

The generator model $\mathcal{G}$ in GANs is responsible for refining the representation of low-quality generated resumes and aims to improve it to resemble a high-quality resume representation. The generator's objective is to generate refined representations that can fool the discriminator $\mathcal{D}$ into classifying them as high-quality resumes.
To train the generator model, we propose to simultaneously minimize the generator loss measured by the discriminator model $\mathcal{D}$, i.e.,
\begin{equation}\label{loss_g}
 \min_{\Theta_{\mathcal{G}}}\mathcal{L}_\mathcal{G}  = \mathbb{E}_{c_i \sim \hat{T}_\mathcal{C}^{\uparrow} }[ 1 - \log \mathcal{D}(\mathcal{G}(x_{c_i}))]
\end{equation}
where $\Theta_{\mathcal{G}}$ denote the parameters in the generator model $\mathcal{G}$.

By iteratively training the generator and discriminator in a competitive manner, the discriminator becomes increasingly skilled at distinguishing between real and generated samples, while the generator becomes more adept at producing samples that are harder to distinguish from real data. This adversarial training process drives both the generator and discriminator to improve, ultimately leading to the generation of high-quality samples that closely resemble the real data distribution.

The generator's role is crucial in achieving more accurate and effective job recommendation results by iteratively refining the low-quality resumes and aligning them with high-quality resumes within our proposed LGIR (LLM-based GANs Interactive Recommendation) framework.

\subsection{Multi-objective Learning for Recommendation}

To explore the generation of "high-quality" resume representations for improved recommendation, we utilize the Classifier $\mathcal{C}$ and Generator $\mathcal{G}$ to obtain aligned resume representations, denoted as $z_{c_i}$, for all users, regardless of whether they are few-shot users or many-shot users, i.e.,

\begin{equation}
z_{c_i} = \begin{cases} x_{c_i}  ,  &\text{if } \mathcal{C}(x)\ge 0.5;\\
\mathcal{G}(x_{c_i})  ,  &\text{if } \mathcal{C}(x) < 0.5  \\
\end{cases}
\end{equation}

To predict users' preference on jobs, we propose a deep model to capture the non-linear and complex relationship between the user $c_i$ and the job $j_k$, i.e.,

\begin{equation}\label{loss_rec}
\hat{R}_{i,k} = g(c_i,j_k)= W^p \cdot [z_{c_i}+x_{j_k};z_{c_i}-x_{j_k};z_{c_i}\odot x_{j_k}]
\end{equation}
where $\odot$ denotes the element-wise product, $W^p \in \mathbb{R}^{1\times 3\cdot d_e}$ maps to a score or
probability of $j_k$ that user $c_i$ will engage.

For recommendation target, we adopt the pairwise loss to define the recommendation objective function as follows,

\begin{equation}
\mathcal{L}_{rec} = \max\limits_{\Theta} \sum_{(i,j_1,j_2)\in D}\log
\sigma(\hat{R}_{i,j_1} - \hat{R}_{i,j_2})  - \lambda ||\Theta||^2
\end{equation}
where the train set $D = \{(c_i,j_{1},j_{2})\}$ means that user $c_k$ gave positive feedback to job $j_1$ (i.e., ${R}_{i,j_1}=1$) instead of job $j_2$ (i.e., ${R}_{i,j_2}=1$).
The $\Theta$ denotes all parameters need to be learned
in the proposed model and $\lambda$ is the regularization coefficient of L2 norm $||\cdot||^2$.
The objective function shows that the job $j_1$ with
positive feedback should have a higher score than the job $j_2$. We summary the whole train process of the proposed model as in algorithm 1.

\begin{algorithm}[tb]
\caption{Algorithm of the proposed method. }
\label{alg:algorithm}
\textbf{Input}: User set $\mathcal{C}$ and job set $\mathcal{J}$, their associated
text document $T$ and interactions $\mathcal{R}$, threshold hyper-parameters $\kappa_1$ and $\kappa_2$, learning rate $\eta$.
\begin{algorithmic}[1] 
\STATE Initialize all parameters $\Theta$ need to be learned.
\WHILE{stop condition is not reached}
\STATE \# Training the classifier $\mathcal{C}$.
\FOR{number of training iterations}
 \STATE  Sample an instance $(c_i,\hat{y}_{c_i})\sim  T_\mathcal{C}$. Update $\mathcal{C}$ by ascending along its stochastic gradient according to Equation (\ref{loss_c}):  $\Theta_{\mathcal{C}} \gets \Theta_{\mathcal{C}} - \eta\cdot \frac{\partial \mathcal{L}_\mathcal{C}}{\Theta} $;
\ENDFOR
\STATE \# Training the discriminator model $\mathcal{D}$:
\FOR{number of training iterations}
\STATE  Adjust set $\hat{T}_\mathcal{C}^{\uparrow}$ and set $\hat{T}_\mathcal{C}^{\downarrow}$ according to classifier $\mathcal{C}$.
 \STATE  Sample a pair  $({c_{i_1}},{c_{i_2}})\sim  \hat{T}_\mathcal{C}^{\uparrow}\times \hat{T}_\mathcal{C}^{\downarrow}$. Update $\mathcal{D}$ by ascending along its stochastic gradient according to Equation (\ref{loss_d}):  $\Theta_{\mathcal{D}} \gets \Theta_{\mathcal{D}} - \eta\cdot \frac{\partial \mathcal{L}_\mathcal{D}}{\Theta} $;
\ENDFOR
\STATE \# Training the generator model $\mathcal{G}$ :
\FOR{number of training iterations}
 \STATE  Sample an instance  ${c_{i}}\sim  \hat{T}_\mathcal{C}^{\uparrow}$. Update $\mathcal{G}$ by ascending along its stochastic gradient according to Equation (\ref{loss_g}):  $\Theta_{\mathcal{G}} \gets \Theta_{\mathcal{G}} - \eta\cdot \frac{\partial \mathcal{L}_\mathcal{C}}{\Theta} $;
\ENDFOR
\STATE \# Training the recommendation target:
\FOR{number of training iterations}
 \STATE  Sample a triplet  $(c_i,j_{1},j_{2}) \in D$. Update all parameters by ascending along its stochastic gradient according to Equation (\ref{loss_rec}):  $\Theta \gets \Theta - \eta\cdot \frac{\partial \mathcal{L}_{rec}}{\Theta} $;
\ENDFOR
\ENDWHILE
\end{algorithmic}
\end{algorithm}

\section{Experiment} \label{Sec_Experiments}
In this section, we aim to evaluate the performance and
effectiveness of the proposed method. Specifically, we conduct
several experiments to study the following research questions:

\begin{itemize}
\item \textbf{RQ1}: Whether the proposed method LGIR outperforms state-of-the-art methods for job recommendation? 

\item \textbf{RQ2}: Whether the proposed method benefits from inferring users' implicit characteristics from their behaviors for more accurate and meaningful resume generation?

\item \textbf{RQ3}: Whether the proposed LGIR framework benefits from aligning the generated resumes of few-shot users with high-quality representations?

\item \textbf{RQ4}: How do different configurations of the key hyper-parameters impact the performance of LGIR?

\end{itemize}

\subsection{Experimental Setup}
 \subsubsection{Datasets}

We evaluated the proposed method on three real-world data sets, which was provided by a popular online recruiting platform. These data sets were collected from 106 days of real online logs for job recommendation in the designer, sales, and technology industries, respectively. These data sets contained the rich interaction between users and employers. In addition, these data sets also contained text document information, which were the resumes of the users and the descriptions of job positions. The characteristics of these data sets are summarized in Table \ref{Table_es}.

\begin{table}[htb]
\caption{Statistics  of the experimental data sets} \centering
\begin{tabular}{p{1.3cm}p{1cm}p{1cm}p{1.4cm}l}
\hline\noalign{\smallskip}
Data set &\#Users &\#Items&\#Interaction  \\
\noalign{\smallskip}\hline\noalign{\smallskip}
Designs &  12,290& 9,143& 166,270 \\
Sales &  15,854&  12,772 &  145,066 \\ 
Tech &  56,634& 48,090&925,193 \\
\noalign{\smallskip} \hline
\end{tabular} \label{Table_es}
\end{table}

\subsubsection{Evaluation Methodology and Metrics}

We spitted the interaction records into training, validation, and test sets equally for the model training, validation, and testing phases.  Experimental results were recorded as the average of five runs with different random initialization of model parameters.

To evaluate the performance of different methods,
we adopted three widely used evaluation metrics for top-$n$
recommendation \cite{Zhao22},  mean average precision ($map@n$), normalized discounted cumulative gain ($ndcg@n$) and mean reciprocal rank ($mrr$), where $n$ was set as 5 empirically. We sampled $20$ negative instances for each positive instance from users' interacted and non-interacted records.

\renewcommand{\arraystretch}{1.4}
\begin{table*}[tp]
  \centering
  \fontsize{9}{11}\selectfont
  \caption{Performance of the proposed and baseline methods for job recommendation. \\ $*$ indicates that the improvements are significant at the level of 0.01  with paired t-test.}
  \label{tab:RSComparison}
\begin{tabular}{|l|lll|lll|lll|}
\hline
                  & \multicolumn{3}{c|}{designs}                                                                  & \multicolumn{3}{c|}{sales}                                                                    & \multicolumn{3}{c|}{tech}                                                                     \\ \hline
                  & \multicolumn{1}{l|}{map@5}           & \multicolumn{1}{l|}{ndcg@5}          & mrr             & \multicolumn{1}{l|}{map@5}           & \multicolumn{1}{l|}{ndcg@5}          & mrr             & \multicolumn{1}{l|}{map@5}           & \multicolumn{1}{l|}{ndcg@5}          & mrr             \\ \hline
SGPT-BE           & \multicolumn{1}{l|}{0.0712}          & \multicolumn{1}{l|}{0.1140}          & 0.2128          & \multicolumn{1}{l|}{0.0526}          & \multicolumn{1}{l|}{0.0932}          & 0.1726          & \multicolumn{1}{l|}{0.1464}          & \multicolumn{1}{l|}{0.2092}          & 0.3344          \\ \hline
SGPT-ST           & \multicolumn{1}{l|}{0.0694}          & \multicolumn{1}{l|}{0.1107}          & 0.2077          & \multicolumn{1}{l|}{0.0519}          & \multicolumn{1}{l|}{0.0926}          & 0.1714          & \multicolumn{1}{l|}{0.1422}          & \multicolumn{1}{l|}{0.2025}          & 0.3289          \\ \hline
SGPT-ST + SRC & \multicolumn{1}{l|}{0.0727}          & \multicolumn{1}{l|}{0.1177}          & 0.2185          & \multicolumn{1}{l|}{0.0511}          & \multicolumn{1}{l|}{0.0925}          & 0.1719          & \multicolumn{1}{l|}{0.1541}          & \multicolumn{1}{l|}{0.2194}          & 0.3442          \\ \hline
BPJFNN            & \multicolumn{1}{l|}{0.1415}          & \multicolumn{1}{l|}{0.2156}          & 0.3436          & \multicolumn{1}{l|}{0.1138}          & \multicolumn{1}{l|}{0.2038}          & 0.3030          & \multicolumn{1}{l|}{0.2018}          & \multicolumn{1}{l|}{0.2948}          & 0.4704          \\ \hline
MF                & \multicolumn{1}{l|}{0.1914}          & \multicolumn{1}{l|}{0.2913}          & 0.4557          & \multicolumn{1}{l|}{0.0887}          & \multicolumn{1}{l|}{0.1628}          & 0.2789          & \multicolumn{1}{l|}{0.4359}          & \multicolumn{1}{l|}{0.6054}          & 0.7555          \\ \hline
NCF               & \multicolumn{1}{l|}{0.2071}          & \multicolumn{1}{l|}{0.3230}          & 0.4944          & \multicolumn{1}{l|}{0.1463}          & \multicolumn{1}{l|}{0.2670}          & 0.3941          & \multicolumn{1}{l|}{0.4105}          & \multicolumn{1}{l|}{0.5706}          & 0.7414          \\ \hline
PJFFF             & \multicolumn{1}{l|}{0.1182}          & \multicolumn{1}{l|}{0.1855}          & 0.3299          & \multicolumn{1}{l|}{0.0690}          & \multicolumn{1}{l|}{0.1255}          & 0.2199          & \multicolumn{1}{l|}{0.2802}          & \multicolumn{1}{l|}{0.4040}          & 0.6127          \\ \hline
SHPJF             & \multicolumn{1}{l|}{0.1862}          & \multicolumn{1}{l|}{0.2875}          & 0.4531          & \multicolumn{1}{l|}{0.1334}          & \multicolumn{1}{l|}{0.2436}          & 0.3705          & \multicolumn{1}{l|}{0.3710}          & \multicolumn{1}{l|}{0.5189}          & 0.7016          \\ \hline
LightGCN        & \multicolumn{1}{l|}{{\ul 0.2664}}    & \multicolumn{1}{l|}{{\ul 0.4218}}    & {\ul 0.5955}    & \multicolumn{1}{l|}{{\ul 0.1629}}    & \multicolumn{1}{l|}{{\ul 0.2980}}    & {\ul 0.4271}    & \multicolumn{1}{l|}{0.4676}          & \multicolumn{1}{l|}{0.6591}          & 0.8093          \\ \hline
LightGCN+SRC      & \multicolumn{1}{l|}{0.2649}          & \multicolumn{1}{l|}{0.4189}          & 0.5926          & \multicolumn{1}{l|}{0.1611}          & \multicolumn{1}{l|}{0.2939}          & 0.4204          & \multicolumn{1}{l|}{{\ul 0.4719}}    & \multicolumn{1}{l|}{{\ul 0.6661}}    & {\ul 0.8146}    \\ \hline

\textbf{LGIR(ours)}     & \multicolumn{1}{l|}{\textbf{0.2887*}} & \multicolumn{1}{l|}{\textbf{0.4622*}} & \textbf{0.6319*} & \multicolumn{1}{l|}{\textbf{0.1751*}} & \multicolumn{1}{l|}{\textbf{0.3225*}} & \textbf{0.4548*} & \multicolumn{1}{l|}{\textbf{0.5086*}} & \multicolumn{1}{l|}{\textbf{0.7191*}} & \textbf{0.8434*} \\ \hline
imprv             & \multicolumn{1}{l|}{8.38\%}          & \multicolumn{1}{l|}{9.57\%}          & 6.11\%          & \multicolumn{1}{l|}{7.50\%}          & \multicolumn{1}{l|}{8.22\%}          & 6.49\%          & \multicolumn{1}{l|}{7.78\%}          & \multicolumn{1}{l|}{7.96\%}          & 3.54\%          \\ \hline
\end{tabular}
\end{table*}

\subsubsection{Baselines}

We took the following state-of-the-art methods as the baselines, including content-based methods \cite{muennighoff2022sgpt,qin2018enhancing}, collaborative filtering based methods \cite{BPR,he2017neural,neve2019latent}, hybrid methods \cite{he2020lightgcn,Jiang2020,hou2022leveraging}, and LLMs based method \cite{muennighoff2022sgpt}.

\begin{itemize}
\item[-]\textbf{SGPT-BE\cite{muennighoff2022sgpt}}: It applies GPT models as \underline{B}i-\underline{E}ncoders to asymmetric search, which produces semantically meaningful sentence embeddings by contrastive fine-tuning of only bias tensors and position-weighted mean pooling.
\item[-]\textbf{SGPT-ST\cite{reimers-2019-sentence-bert}}: It use SGPT with \underline{S}entence \underline{T}ransformers as encoders to asymmetric search. 
\item[-]\textbf{SGPT-ST+SRC\cite{muennighoff2022sgpt}}: We adopt the \underline{S}imple \underline{R}esume \underline{C}ompletion strategy for resume completion,  and we then adopt SGPT-ST for  asymmetric search. 
\item[-]\textbf{BPJFNN \cite{qin2018enhancing}}: It adopts a hierarchical attentional RNN model to learn the word-level semantic representations of resumes and job descriptions.
\item[-]\textbf{MF \cite{koren2009matrix}}: It learns low-dimensional representations of users and items by reconstructing their interaction matrix based on the point loss.
\item[-]\textbf{NCF \cite{he2017ncf}}: It enhances collaborative filtering with deep neural networks, which adopt an MLP  to explore the non-linear interaction between user and item.
\item[-]\textbf{PJFFF \cite{Jiang2020}}: It fuses the representations for the explicit and implicit intentions of users and employers by the historical application records.
\item[-]\textbf{SHPJF \cite{hou2022leveraging}}: It utilizes both text content from jobs or resumes and search histories of users for recommendation. 
\item[-]\textbf{LightGCN\cite{he2020lightgcn}}: It simplifies the vanilla GCN's implementation to make it concise for collaborative filtering. Furthermore, we enhance the LightGCN with text information based on BERT model for job recommendation.
\item[-]\textbf{LightGCN+SRC\cite{he2020lightgcn}}:  We enhance the LightGCN with generated texts by the \underline{S}imple \underline{R}esume \underline{C}ompletion strategy for job recommendation.
\end{itemize}

\begin{table}[tp]
  \centering
  \fontsize{9}{11}\selectfont
  \caption{Performance of the variants for ablation studies. }
  \label{table_ablation}
\begin{tabular}{|c|l|l|l|l|}
\hline
\multicolumn{1}{|l|}{Dataset} & Method        & map@5           & ndcg@5          & mrr             \\ \hline
\multirow{4}{*}{designs}      & BASE          & 0.2627          & 0.4128          & 0.5829          \\ \cline{2-5} 
                              & SRC           & 0.2601          & 0.4076          & 0.5781          \\ \cline{2-5} 
                              & IRC           & 0.2859          & 0.4560          & 0.6220          \\ \cline{2-5} 
                              & LGIR & \textbf{0.2887} & \textbf{0.4622} & \textbf{0.6319} \\ \hline
\multirow{4}{*}{sales}        & BASE          & 0.1617          & 0.2945          & 0.4250          \\ \cline{2-5} 
                              & SRC           & 0.1652          & 0.3031          & 0.4331          \\ \cline{2-5} 
                              & IRC           & 0.1671          & 0.3065          & 0.4359          \\ \cline{2-5} 
                              & LGIR & \textbf{0.1751} & \textbf{0.3225} & \textbf{0.4548} \\ \hline
\multirow{4}{*}{tech}         & BASE          & 0.4994          & 0.7088          & 0.8374          \\ \cline{2-5} 
                              & SRC           & 0.5048          & 0.7148          & 0.8435          \\ \cline{2-5} 
                              & IRC           & 0.5056          & 0.7153          & 0.8400          \\ \cline{2-5} 
                              & LGIR & \textbf{0.5086} & \textbf{0.7191} & \textbf{0.8434} \\ \hline
\end{tabular}
\end{table}

\subsubsection{Implementation Details}
We adopted the ChatGLM-6B \cite{du2022glm} as the LLM model in this paper. For a fair comparison, all methods were optimized by the AdamW optimizer with the same latent space dimension (i.e., $64$), batch size (i.e., $1024$), learning rate (i.e., $5\times10^{-5}$), and regularization coefficient (i.e., $10^{-4}$). We set $d=768$, $d_{e'}=128$, $d_e=64$, and $d_c=d_s=d_g=256$ for the proposed method.
We carefully searched other special hyper-parameters for best performance. Early stopping was used with the patience of $50$ epochs. 

\begin{figure*} \centering
 \includegraphics[width=1.0\textwidth]{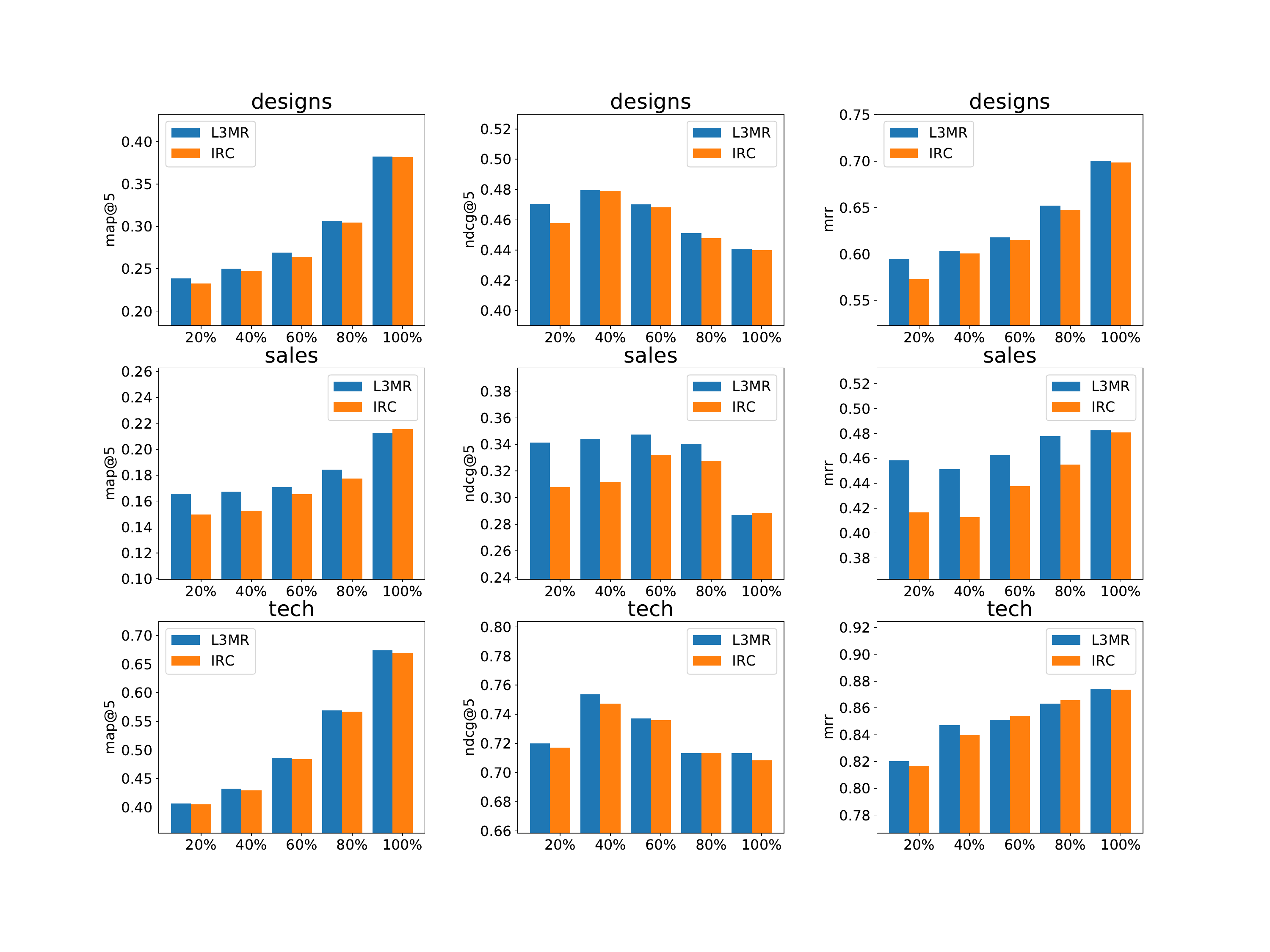}
\caption{{Performance comparison of LGIR and the variant IRC for few-shot analysis.}} \label{fig_fewshot}
\end{figure*}

\begin{figure*} \centering
 \includegraphics[width=0.9\textwidth]{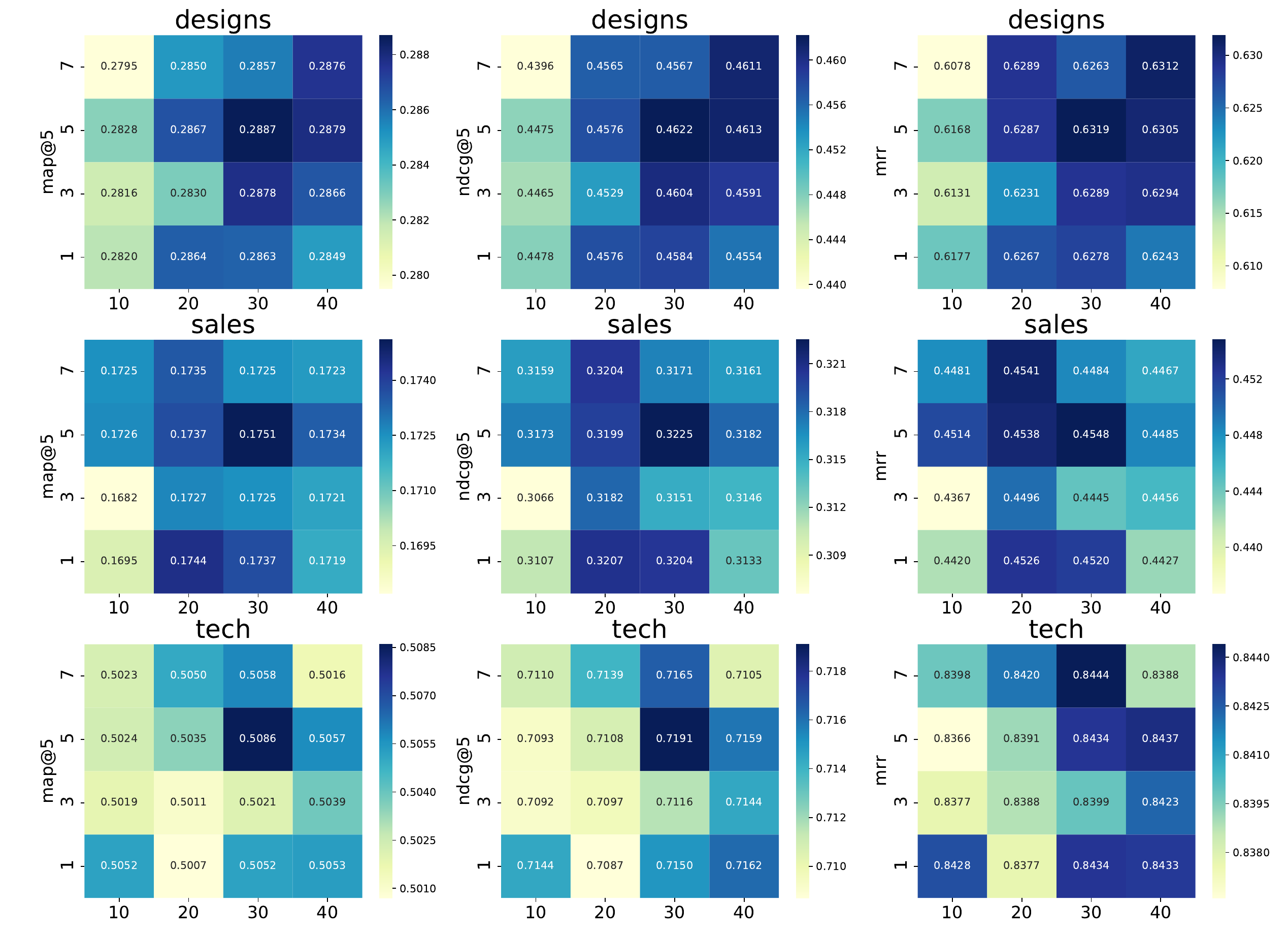}
\caption{{Effects of threshold parameters $\kappa_1 \in \{10, 20, 30,40\}$ and $\kappa_2 \in \{1, 3, 5,7\}$  for  the few-shot users and  many-shot users selection.}} \label{heatmap}
\end{figure*}

\subsection{Model Comparison (RQ1)}

Table \ref{tab:RSComparison} shows the performance of different methods for job recommendation. To make the table more notable, we mark the top-2 results for each data set. From the experimental results, we can get the following conclusions:
\begin{itemize}
\item  Firstly, the proposed method LGIR consistently outperforms all baseline methods in all cases, demonstrating the effectiveness of the proposed method. Specifically, our model can
on average improve the best baseline by 8.02\%, 7.40\%, and 6.42\% relatively on designs, sales, and tech datasets, respectively.
\item Secondly, it is noteworthy that the LLMs methods (such as SGPT-BE, SGPT-ST, and SGPT-ST+SRC) that do not take into account users' interaction behaviors demonstrate poor performance compared to other approaches. This observation strongly suggests that relying solely on textual descriptions of users and jobs may not be an effective strategy for job recommendation. The poor performance of these methods can be attributed to the inherent limitations of using textual descriptions, such as the presence of default, meaningless, or erroneous information in the descriptions of users and jobs.
\item Thirdly, some hybrid methods (such as PJFFF and SHPJF), whose embeddings of users and jobs highly rely on the text content, do not perform well in most cases. This could be attributed to the requirement for texts to be structured and unabridged, whereas in our scenario, users on the platform have different text organization habits.
\item The GCN-based methods (e.g., LightGCN and LightGCN + SRC), whose embeddings of users and jobs rely on both their preference encoding and text content, achieve the best performance across baselines, indicating that utilizing both interactions and text descriptions is important for job recommendation. 
\item Finally, the simple resume completion (SRC) strategy  shows limited improvement (e.g., LGCN vs. LGCN + SRC) in most cases, which indicates that simply leveraging LLMs to enhance job recommendation is not a one-size-fits-all solution because LLMs have the limitation of fabricated and hallucinated generation. This motivates us to extract accurate and valuable information beyond users' self-description to help the LLMs to better profile users for resume completion.
\end{itemize}

\subsection{Ablation Study (RQ2\&3)}
To evaluate the effectiveness of the module design in the proposed method LGIR, we compare it to several special cases which include:

\begin{itemize}

 \item[-]\textbf{BASE}: This method is a two-tower text matching model for job recommendation, which utilize the original self-description provided by users. 
 \item[-]\textbf{SRC}:  This method uses the generated resume of users with the simple resume completion (SRC) strategy without GANs-based learning for job recommendation.
 \item[-]\textbf{IRC}: This method uses the generated resume of users with the interactive resume completion (IRC) strategy for job recommendation. We eliminate the GANs-based learning for aligning the unpaired user resumes. 
 \item[-]\textbf{LGIR}: This method is the proposed method in this paper. Specifically, LGIR contains the interactive resume completion (IRC) strategy and the GANs-based learning for job recommendation.
\end{itemize}

Table \ref{table_ablation} shows the performance of the proposed method and ablation methods,
i.e. LGIR, BASE, SRC, and IRC. From the experimental results, we can get the following conclusions:

\begin{itemize}

\item RQ2: The variant SRC which completes the resume of users with LLMs show limited improvement compared to the variant BASE in most cases, which indicates that simply leveraging LLMs to enhance job recommendation is not a one-size-fits-all solution. To alleviate the limitation of the fabricated and hallucinated generation of LLM, we propose an interactive resume completion (IRC) strategy for high-quality resume completion. Specifically, the variant IRC shows significant improvement to the variant BASE and the variant IRC, which indicates the necessity of inferring users' implicit characteristics based on their behaviors for more accurate and meaningful resume generation.

\item RQ3: The proposed method LGIR shows significant improvement to the variant IRC on all data sets, which benefits from the GANs-based learning to align the generated resumes of few-shot users with the high-quality representations. The in-depth analysis of how GANs refine the generated resumes of users is presented in the following few-shot analysis.
\end{itemize}

\begin{figure*} \centering
 \includegraphics[width=0.95\textwidth]{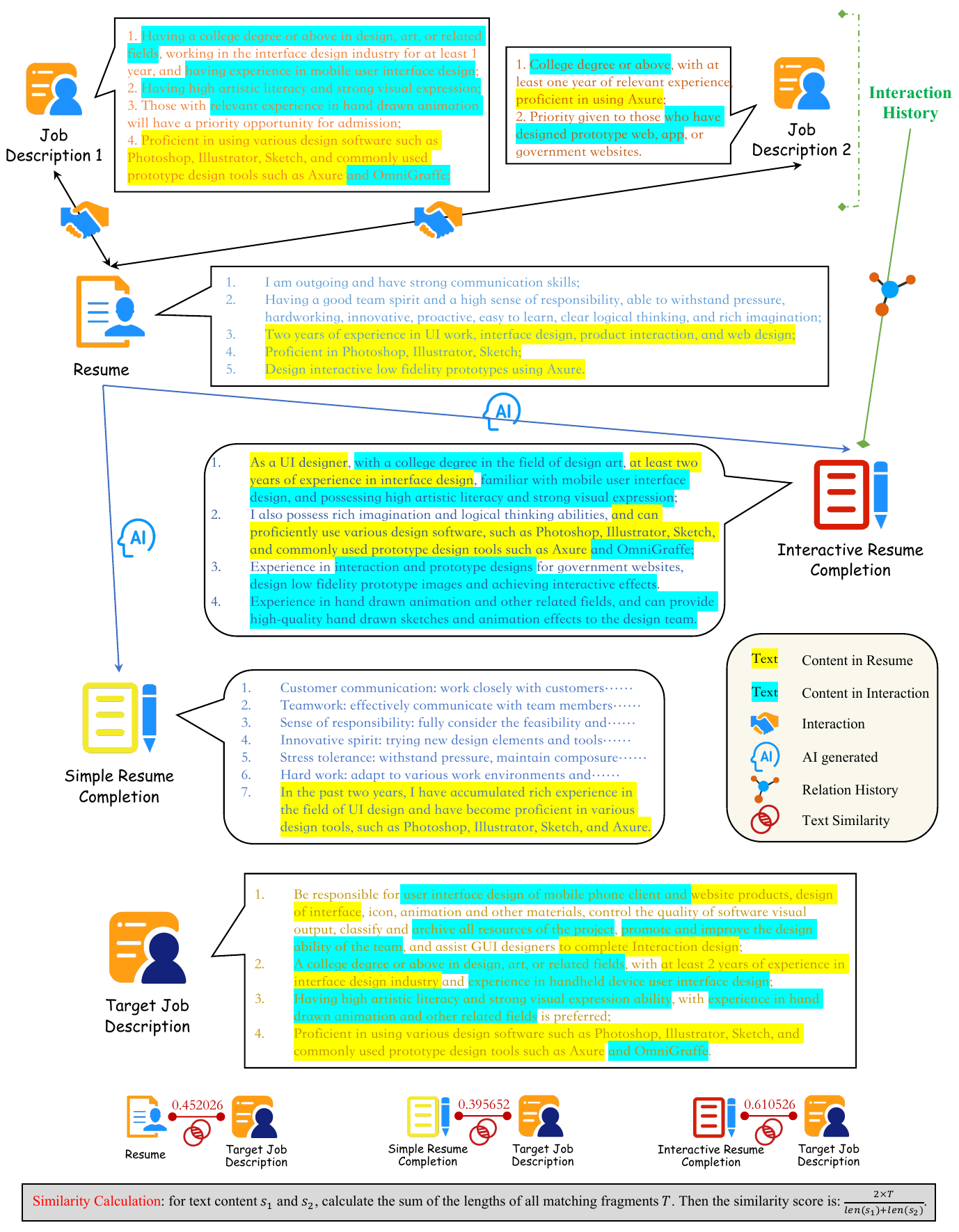}
\caption{{A real recruitment scenario where users have two historical interactions. The process explains how the model successfully integrates pertinent information from user resumes and interactive job descriptions, resulting in generated resumes that better reflect the user's abilities. And it has been proven through calculating text similarity that our method can numerically improve the success rate of recommendation system matching.}} \label{case_study}
\end{figure*}

\subsection{Few-shot Analysis (RQ3)}

The ablation study shows that the proposed method LGIR benefits from the GANs to align the generated resumes of few-shot users with high-quality resumes. It is interesting to investigate whether LGIR can capture our claims at the case level. To this end, we conduct a few-shot analysis that compares the proposed method LGIR with the variant \textbf{IRC} among different shots levels. Specifically, we divided all the users into five groups based on the shot number of their interactions, with the overall number of users in each group remaining equal. For example, the group $40\%$ denotes the user set that falls within the $20\% - 40\%$ ranking range based on the shot number of interactions. Then, we compare the recommendation performance of LGIR (with GANs) with the variant IRC (without GANs) on these overall five groups and report the results in Fig.\ref{fig_fewshot}.  On the one hand, the performance of LGIR is better than the variant IRC in most cases, which indicates the effectiveness of the proposed GANs-based learning scheme. On the other hand, the proposed method LGIR exhibits greater improvement than the variant \textbf{IRC} in the group with fewer shots (i.e., interactions), indicating that GANs-based learning can align the generated resumes of few-shot users with the high-quality resumes of users who have the rich interaction records.  Therefore, the proposed method can effectively alleviate the few-shot problems that limit the quality of resume generation.

\subsection{Hyper-parameter analysis (RQ4)}
We also evaluate the effects of different settings of the key hyper-parameters on the performance of the proposed method, including the threshold parameters $\kappa_1$ and $\kappa_2$ which are set to select the few-shot users and many-shot users for classifier and GANs learning.
The threshold parameters $\kappa_1$ is chosen from $\kappa_1 \in \{10, 20, 30,40\}$, while the threshold parameters $\kappa_2$ is chosen from $\kappa_2 \in \{1, 3, 5,7\}$. 

Figure~\ref{heatmap} shows the performance of the proposed method with different settings of the threshold parameters $\kappa_1$ and $\kappa_2$. It shows that the proposed method achieves the best performance with $\kappa_1 = 30$ and $\kappa_2=5$ in all cases. In other words, we can select the users with more than 30 interaction records as the many-shot users, based on whose self-description and behaviors we can generate high-quality resumes with LLMs. Adversely, we select the users with less than 5 interaction records as the few-shot users for training the classifier.

\subsection{Case Study}

With the impressive generation capabilities of large language models, there is an opportunity to delve deeper into the outputs they produce at each stage and how they can assist LGIR in achieving state-of-the-art results. To illustrate the effectiveness of this approach, we present a real recruitment scenario as an example, as depicted in Fig.\ref{case_study}.

The upper section of the figure showcases the user's resume and two job descriptions that the user has previously interacted with, representing the knowledge base of the user's interaction history. In the middle section of the image, we demonstrate the outcomes of resume completion using two approaches: LLMs alone (referred to as Simple Resume Completion) and LLMs guided by interactive history (referred to as Interactive Resume Completion). Following that, we provide the description of the target position matched by the user. The content in the user's resume that shares a similar meaning with the target job description is highlighted in yellow, while the content that does not appear in the user's resume but exists in the job description of the interaction history is highlighted in blue.

It is evident that the user's interaction history contains valuable common competency information relevant to the target job (highlighted in blue), which is absent in the user's own resume. If we solely rely on the user's resume and allow the LLM to be directly updated (as shown in Simple Resume Completion), it fails to incorporate effective information related to the target job. Due to the potential hallucinations and nonsensical characteristics of LLMs, there is a tendency to generate a plethora of invalid content, thereby reducing the proportion of effective information in the resume.
In contrast, the LLMs guided by the interactive historical knowledge base successfully integrate pertinent information from user resumes and interactive job descriptions, resulting in generated resumes that better reflect the user's abilities, even those they may not have fully expressed or been aware of. This approach effectively enhances the success rate of the recommendation system's matching process.

Furthermore, in addition to visualization, we have also quantified this feature in the bottom section of the image. By calculating the pairwise similarity between texts, we discovered that directly matching the user's resume with the target job description yields a similarity score of approximately 0.45. However, when relying solely on LLMs to expand resumes, the introduction of numerous hallucinations leads to a decrease in similarity, approximately 0.39. On the other hand, through interactive resume completion, the similarity between texts significantly improves to 0.61, representing an around remarkable 35\% enhancement. 

Therefore, exploiting the interactive behaviors of users help LLMs to accurately capture users’ skills and preferences, which contribute to the effectiveness of job recommendation.

\section*{Conclusion}

In this paper,  we propose an LLM-based GANs Interactive
Recommendation (LGIR) method for job recommendation. To
alleviate the limitation of the fabricated generation of LLMs,
we infer users’ implicit characteristics from their behaviors for more accurate and meaningful resume completion. To alleviate the few-shot problems when resume generation, we propose the transfer representation learning with GANs to refine low-quality resumes. The proposed method outperforms state-of-the-art baseline methods, which demonstrates the superiority of utilizing LLMs with interactive resume completion and alignment for job recommendation. The ablation study shows the importance of each component of LGIR, and the case study further illustrates the superiority of LGIR in capturing users’ skills and preferences by LLMs.

\bibliographystyle{IEEEtran}
\bibliography{sample}

\end{document}